\documentclass[aps,twocolumn,showpacs,superscriptaddress]{revtex4}
\usepackage{psfig}
\usepackage{graphicx}
\usepackage{epsf}
\usepackage{graphics,epsf}
\newcommand{\bea}{\begin{eqnarray}}
\newcommand{\beq}{\begin{equation}}
\newcommand{\eea}{\end{eqnarray}}
\newcommand{\eeq}{\end{equation}}
\begin{document}
\bibliographystyle{revtex}
\title{Backbending and $\gamma$ vibrations}
\vspace{0.5cm}
\author{J. Kvasil}
\affiliation{Institute of Particle and Nuclear Physics, Charles
University, V.Hole\v sovi\v ck\'ach 2, CZ-18000 Praha 8, Czech Republic}
\author{R. G.~Nazmitdinov}
\affiliation{Departament de F{\'\i}sica,
Universitat de les Illes Balears, E-07122 Palma de Mallorca, Spain}
\affiliation{Bogoliubov Laboratory of Theoretical Physics,
Joint Institute for Nuclear Research, 141980 Dubna, Russia}

\begin{abstract}
 We propose that the backbending phenomenon
 can be explained as a result of the disappearance of collective
 $\gamma$-vibrational mode  in the rotating frame.
 Using a cranking+random phase approximation approach
 for the modified Nilsson potential + monopole pairing forces,
 we show that this mechanism is responsible for the backbending
 in $^{156}$Dy and $^{158}$Er, and
 obtain a good agreement between theoretical and experimental results.
\end{abstract}
\pacs{21.10.Re,21.60.Jz,27.70.+q}
\maketitle

There is a general persuasion that the backbending is caused by
the rotational alignment of angular momenta of a nucleon pair
occupying a high-j intruder orbital near the Fermi surface. It is
assumed that the alignment breaks a singlet Cooper pairing in this
pair and leads to a sudden increase of the kinematical moment of
inertia ${\cal J}^{(1)}=I/\Omega$ along the yrast level sequence
as a function of a rotational frequency $\Omega$ (cf
Ref.\onlinecite{RS80}). Indeed, in many cases this single-particle
(quasiparticle) mechanism is supported by microscopic analysis in
terms of various cranking Hartree-Fock-Bogoliubov calculations (cf
\cite{tan,eg,fr}). It should be noted, however, that the role of
vibrational (collective) excitations in the backbending has {\it
never} been studied.

It is well known that mean field description of finite Fermi
systems could break spontaneously one of the symmetries of the
exact Hamiltonian, the so-called spontaneously symmetry breaking (SSB)
phenomenon. In fact, the mean field description of the backbending
corresponds to the first-order phase transition. However, this
concept is appropriate only in the limit of infinite number of
particles. Obviously, for finite systems 
quantum fluctuations, beyond the mean field
approach, are quite important.
The random phase approximation (RPA) being  an efficient tool to
study these quantum fluctuations (vibrational and rotational
excitations) provides also a consistent way to
restore broken symmetries. Moreover, it separates the collective
excitations associated with each broken symmetry as a spurious RPA
mode and fixes the corresponding inertial parameter. In this paper
we present the first self-consistent quantitative treatment of
these ideas for rotating nuclei \cite{spain}. We demonstrate that
the backbending in $^{156}$Dy and $^{158}$Er can be explained as a
result of vanishing one of the quadrupole vibrational modes in the
rotating frame at a critical rotational frequency. 
Consequently, collective motion associated with this mode should describe rotational
states of the nonaxial rotating system.

 The practical application of the RPA for nonseparable effective forces such as
the Gogny or Skyrme interactions in rotating nuclei requires too
large configuration space and still is not-available.
The RPA with separable
multipole-multipole interaction based upon phenomenological
cranking Nilsson or Saxon-Woods potentials with pairing forces
gives a sufficiently good description of low-lying collective
excitations in rotating nuclei (cf \cite{KN,NM96}).
Following this approach, which hereafter is called CRPA, we start
with the cranking shell model (CSM)  Hamiltonian in the form
\beq
\label{h1}
H_{\Omega} = H - \sum_{\tau =N,P} \lambda_{\tau}\hat N_{\tau}-\,
\Omega \hat J_x+\, H_{\rm int}.
\eeq
The term  $H=H_N\,+\, H_{\rm add}$ contains a Nilsson Hamiltonian
$H_N$ with three different
oscillator frequencies $\omega_i^2=\omega_0(\beta,\gamma)^2
\Bigl[ 1 - 2 \beta \sqrt{\frac{5}{4 \pi}} cos(\gamma -
\frac{2 \pi}{3}i) \Bigr] \quad (i=1,2,3)$
that determine quadrupole deformation
parameters $\gamma$ and $\beta$ (cf \cite{NR}).
The frequencies are subject to the volume conservation constraint
$\omega_x\omega_y\omega_z=\omega_{00}^3$
($\hbar\omega_{00}=41A^{-\frac{1}{3}}$ MeV) that models the nuclear
incompressibility.
In the cranking model with the standard Nilsson
potential the value of the moment of inertia is largely overestimated due to
the presence of the velocity dependent $\vec l\,^2$ term. This shortcoming can
be overcome by introducing the additional term
$H_{\rm add}=\sum_i^A h_{\rm add}(i)$ with
\bea
h_{add} =
\Omega \,m\, \omega_{00} \kappa\Bigg{[}
&2&\left(r^2 s_x - x \vec r \cdot \vec s\right)\\
&+&\mu \left(2 r^2 - \frac{\hbar}{m\omega_{00}}
(N+\frac{3}{2})\right)\,l_x \Bigg{]}.\nonumber
\eea
The term restores the local Galilean invariance of the Nilsson potential in
the rotating frame and removes the spurious effects of the $\vec
l\,^2$ term (see for details Ref.\onlinecite{NM96}). Note that this
basic recipe supersedes the fitting procedure of nuclear inertial properties
used, for example, in  Ref.\onlinecite{Frau}.

The interaction is taken in a separable form
\beq
\label{rin}
H_{\rm int}= -\sum_{\tau}G_{\tau} \hat P_{\tau}^{+} \hat P_{\tau} \,
- \frac{1}{2}\kappa_2 \Bigg{[}\sum_{\sigma,m}{\hat Q}_m^{(\sigma)}{}^2+
{\hat M}^{(+)}{}^2\Bigg{]}.
\eeq
Here, $\tau=$ neutrons or protons,
$\hat P^+ = \sum_k c_k^+ c_{\bar k}^+ $ and $c_k^+, c_k$ are creation and
annihilation single-particle operators, respectively.
An index $k$ is labelling a complete set of the oscillator quantum numbers
($|k\rangle = |N l j m \rangle$) and the index ${\bar k}$ denotes the
time-conjugated state \cite{BM1}.
We recall that the K quantum number (a projection of the angular momentum
on the quantization axis) is not conserved
in rotating non-axially deformed systems.
However, the CSM Hamiltonian adheres to
the $D_2$ spatial symmetry with respect to rotation
by the angle $\pi$ around the rotational axis x.
Consequently, all rotational states can be classified by the
quantum number called signature $\sigma=\exp (-i\pi\alpha)$ leading
to selection rules for the total angular momentum
$I=\alpha + 2n$, $n=0,\pm 1, \pm 2 \ldots $.
In particular, in even-even nuclei the lowest rotational (yrast) band
characterized by the positive signature quantum number
$\sigma=1 \,(\alpha = 0)$ consists of even spins only.

The quadrupole operators $\hat Q_m \quad (m=0,1,2)$ are defined by
\beq \label{qo} \hat Q_m^{(\sigma)}=\frac{i^{2 + m
+(\sigma+3)/2}}{\sqrt{2(1+\delta _{m 0})}}
r^2\biggl(Y_{2m}+(-1)^{(\sigma+3)/2}Y_{2 -m}\biggr). \eeq The
monopole interaction is defined by the positive signature operator
$\hat M^{(+)}=r^2 Y_{0}$. Single-particle matrix elements of any
one-body Hermitian operator $\hat F$ $(P,Q,M)$ are determined by
the signature, time-reversal  and Hermitian conjugation properties
of the operator (cf Ref.\onlinecite{KN}). All multipoles are
expressed in terms of the double-stretched coordinates $\bar
q_i=\frac{\omega_i}{\omega_0} q_i$, $(q_i=x,y,z)$. The effective
interaction restores the rotational invariance of the Hamiltonian
$H$ in the limit of the harmonic oscillator potential \cite{SK}.
This is especially important in order to establish a relation
between SSB of the rotating mean field and an appearance of the
corresponding RPA spurious mode (cf \cite{n2}). While a
qualitative discussion about a relation between SSB effects and
RPA spurious modes in {\it nonrotating} nuclei could be found
in literature (cf Ref.\onlinecite{RS80,R}), here we present the first
realistic quantitative attempt to get a thorough insight into this
relation in {\it rotating} nuclei. The self-consistent determination 
of the constants $G$
and $\kappa_2$ will be discussed below.

Using the generalized Bogoliubov transformation for quasiparticles
(for example, for the positive signature quasiparticle we have
$\alpha_i^+=\sum_k{\cal U}_{ki} c_k^+ + {\cal V}_{\bar k i}c_{\bar
k}$) and  the variational principle (see details in
Ref.\onlinecite{KN}), we obtain the Hartree-Bogoliubov (HB)
equations. The HB equations are solved using the "experimental"
values of deformation parameters $\beta$ and $\gamma$ (see
Fig.{\ref{fig1}), obtained in Ref.\onlinecite{Si}, as an input.
These values are extracted from experimental data \cite{lan}.
 For the pairing field
$\Delta_{kl} = -\delta_{kl}G_{\tau} \langle P_{\tau} \rangle =
\delta_{kl} \Delta_{\tau}(\Omega)$
we assume a phenomenological dependence of the pairing gap
$\Delta_{\tau}$ upon the rotational frequency $\Omega$
\beq
\Delta_{\tau}(\Omega) \,=\,
\left \{
\begin{array}{l}
\Delta_{\tau}(0)\,[1-\frac{1}{2} (\frac{\Omega}{\Omega_c})^2 \,] \qquad
for \,\,\,\Omega < \Omega_c \\
\Delta_{\tau}(0)\,\frac{1}{2} (\frac{\Omega_c}{\Omega})^2 \qquad \qquad \,
for \,\,\,\Omega > \Omega_c \\
\end{array},
\right.
\eeq
introduced in Ref.\onlinecite{Wyss}.
Here, $\Omega_c=0.32,0.33$ MeV is a rotational frequency where
the first band crossing (which is approximately
the same for protons and neutrons) occurs for $^{156}$Dy and
$^{158}$Er, respectively.
The values of pairing gaps $\Delta_n(0)=0.857$ MeV,
$\Delta_p(0)=0.879$ MeV (for $^{156}$Dy)
and $\Delta_n(0)=0.874$ MeV, $\Delta_p(0)=0.884$ MeV (for $^{158}$Er)
are obtained from the odd-even mass difference
(see also Ref.\onlinecite{lan})).
The Nilsson parameters $\kappa$ and $\mu$ are taken
from systematic analyses for all deformed nuclei \cite{NR}.
In our calculations we include all shells up to $N=8$
and consider $\Delta N=2$ mixing.
The configuration space exhausted 97\%
of energy weighted sum rule for quadrupole transitions.

\begin{figure}[ht]
\includegraphics[height = 0.15\textheight]{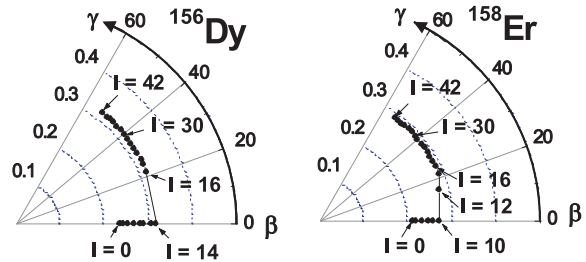}
\caption{(Color online) Equilibrium deformations in $\beta$-$\gamma$ plane
as a function of the angular momentum $I = \langle {\hat J}_x \rangle$
(in units of $\hbar$).}
\label{fig1}
\end{figure}

It is enough to solve the HB equations only for the positive signature
quasiparticle energies $\varepsilon _i $, since
the negative signature eigenvalues  $\varepsilon _{\bar i} $ and
eigenvectors $({\cal U}_{\bar k \bar i}, {\cal V}_{k \bar i})$ are
obtained from the positive ones. Thus, our quasiparticle operators are
defined with respect to the yrast state $| \rangle$ (the lowest HB state
at given $\Omega$) which is our quasiparticle vacuum
$\alpha_i | \rangle=0, \alpha_{\bar i} | \rangle=0$.

In the limit of the harmonic oscillator, at the
self-consistent energy minimum
the expectation values of  the operators
$Q_{0}$ and $Q_{2}^{(+)}$ expressed in the double-stretched coordinates,
Eqs.(\ref{qo}),
are zeros
\beq
\label{cond3}
\langle Q_{0} \rangle = \langle Q_{2}^{(+)} \rangle =0
\eeq
In other words, the self-consistent residual
interaction does not change the mean field equilibrium deformation.
To check the self-consistency of the HB solutions, we
calculate Eqs.(\ref{cond3}) when all terms are included in the
HB equations. First, we would like to stress that
the "experimental" values of $\beta$ and $\gamma$
correspond, indeed, to the minimum of the total mean field (HB) energy.
The variation of the deformation parameters $\beta$ and $\gamma$
around the "experimental" (equilibrium) values results in
the increase of the HB energy.
This minimum becomes very shallow, however, with the increase
of the rotational frequency.
Second, double-stretched quadrupole moments are approximately
zero for all values of the equilibrium deformation parameters.
A small deviation from the equilibrium deformation leads to
a strong deviation  of the double-stretched quadrupole moments from
zero values and correlates with the increase of the HB energy.
The main outcome of the HB calculations  is the onset of the non-axial
deformation around $\hbar\Omega\sim 0.32$ MeV where
the backbending takes place in both nucleus.
While our results are similar to the prediction
of Ref.\onlinecite{Frau} for $N \sim 90$, we 
reproduce better the experimental data. 
Our equilibrium deformations are a result 
of the self-consistent solutions of the HB equations, whereas
the authors of Ref.\onlinecite{Frau} used fixed phenomenological
inertial parameters.

To describe quantum oscillations around mean field solutions
the boson-like operators
$b^{+}_{k\bar{l}}=\alpha^{+}_{k}\alpha^{+}_{\bar{l}},
b^{+}_{kl}=\alpha^{+}_{k}\alpha^{+}_{l},
b^{+}_{\bar{k}\bar{l}}=\alpha^{+}_{\bar{k}}\alpha^{+}_{\bar{l}}$
are used.
The first equality introduces the positive signature boson,
while the other two determine the negative signature ones.
These two-quasiparticle operators are treated in the
quasi-boson approximation (QBA) as an elementary bosons, i.e.,
all commutators between them are approximated by their
expectation values with the uncorrelated HB vacuum \cite{RS80}.
The corresponding commutation relations can be found in Ref.\onlinecite{KN}.
In this approximation any single-particle operator
$\hat{F}$ can be expressed as
$\hat{F} = \langle F \rangle + \hat{F}^{(1)}+\hat{F}^{(2)}$
where the second and third terms are linear and bilinear order terms in
the boson expansion. We recall that in the QBA one includes
all second order terms into the boson Hamiltonian such
that $(\hat{F}-\langle F \rangle)^2=\hat{F}^{(1)}\hat{F}^{(1)}$.
The positive and negative signature boson spaces are not mixed,
since the corresponding operators commute and
$H_{\Omega}=H_\Omega(\sigma=+) + H_\Omega(\sigma=-)$.

The CRPA Hamiltonian is diagonalized by solving the equations of
motion for each signature separately. As a result, we obtain the
following determinant of the secular equations
\begin{equation}
  \label{det}
  {\cal F}(\omega_{\lambda}) =  \det~({\bf R}- \frac{\bf 1}{2 c}~),
\end{equation}
which is the fifth and the second order for the positive and negative
signature, respectively, and $c=\kappa_2$ or $G_{\tau}$.
The matrix elements
$R_{km}(\omega_{\lambda}) =
\sum_\mu q_{k,\mu }q_{m,\mu } C_\mu ^{km}/(E_\mu^2 -\omega_{\lambda}^2)$
involve the coefficients $ C_{\mu}^{km} = E_{\mu }$ or $\omega_{\lambda}$  for
different combinations of matrix elements $q_{k,\mu}$
(see details in Refs.\onlinecite{KN,Al}). The zeros of the function
${\cal F}(\omega_\lambda)=0$
yield the CRPA eigenfrequencies $\omega_\lambda$.
Since the mean field violates the rotational invariance and particle number
conservation law, among the CRPA
eigenfrequencies there exist few spurious solutions.
In this paper we focus our attention on the SSB effects related
to the rotation, since the pairing can be treated in the same way.

Introducing the operator
$\Gamma^{+}=(\hat{J}^{(1)}_y-i\hat{J}^{(1)}_z)/\sqrt{2\langle J_x \rangle}$
such that
\beq
\label{ns}
[H_\Omega(\sigma=-),\Gamma^{+}]=\Omega\Gamma^{+}, \quad
[\Gamma, \Gamma^{+}]=1, \quad \Gamma=(\Gamma^{+})^{+}
\eeq
one is able to separate negative signature vibrational
modes from the "spurious" solution at $\omega_{\lambda}=\Omega$.
 Equations (\ref{ns}) describe a collective 
phonon that creates a collective rotation.
This phonon is related to the symmetry broken  by the external 
rotational field (the cranking term in Eq.(\ref{h1})).

The other spurious solutions are associated with the rotation around the x axis
and the particle number conservation law,
\begin{equation}
\label{ps}
[H_{\Omega}, \hat{J}_x] \,=
\, [H_{\Omega}, \hat N_{\tau}] \,=\,0 .
\end{equation}

The mode associated with the rotation about the x axis
allows one to determine the Thouless-Valatin moment of inertia
${\cal J_{TV}}$ using the positive signature term
of the full Hamiltonian
\beq
\label{tv}
[H_\Omega(\sigma=+), i{\hat \Phi}]=
\frac{\hat J_x}{\cal J_{TV}}, \quad [{\hat \Phi},{\hat J}_x]=i.
\eeq
Here the angle operator ${\hat \Phi}$ is the canonical partner of the
angular momentum operator ${\hat J}_x$.
A similar procedure can be applied for the second spurious mode in 
Eqs.(\ref{ps}) to obtain the mass parameters for neutron or protons~\cite{TV62,MW69}.

It is important to hold a self-consistency
at the CRPA level as well as at the mean field.
In the harmonic oscillator limit
the self-consistent constant,
$ \kappa_2 = \frac{4 \pi}{5} \frac{m \omega_0^2}{ \langle r^2 \rangle}$,
warrants the fulfillment all conservation laws in the CRPA
for rotating nuclei \cite{n2}. Therefore, in our realistic calculations
we define the constants from the requirement
of the fulfillment of the conservation laws, Eqs.(\ref{ps}),
and a separation of the rotational mode from the vibrational ones,
Eqs.(\ref{ns}). It was proved in  Ref.\onlinecite{n2}  for the self-consistent
model, which can be solved exactly, that
the dynamical ${\cal J}_{HB}^{(2)}$ and the Thouless-Valatin
${\cal J}_{TV}$ moments of inertia must coincide,
if one found a {\it self-consistent  mean field
minimum and spurious solutions are separated from the physical
ones}.
Our results (see Fig.\ref{fig2}) demonstrate a good self-consistency
between the mean field and the CRPA calculations, indeed.

\begin{figure}[ht]
\includegraphics[height = 0.25\textheight]{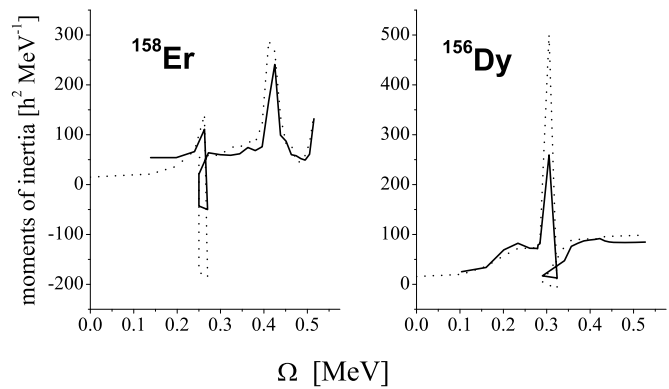}
\caption{The rotational dependence of the
dynamical
${\cal J}_{HB}^{(2)}=-d^2E_{HB}/d\Omega^2=d\langle{\hat J}_x\rangle/d\Omega$
(a dash line) and the Thouless-Valatin ${\cal J}_{TV}$ moments of inertia
(a solid line). Here, $E_{HB}$ is a mean field value of the
full Hamiltonian.}
\label{fig2}
\end{figure}

To analyze the low-lying excited states
we construct the Routhian function for each rotational band
($\nu =$ yrast, $\beta$, $\gamma$, .... )
$R_{\nu}(\Omega) = E_{\nu}(\Omega) - \Omega I_{\nu}(\Omega)$ and
define the experimental excitation energy in the rotating frame
$\omega_{\nu}(\Omega) = R_{\nu}(\Omega) - R_{\rm yr}(\Omega)$ \cite{Na87}.
This energy can be directly compared with the corresponding solutions
$\hbar \omega_{\lambda}$ of the CRPA secular equations.
The experimental Routhians  $R_\nu$ are obtained
from experimental data {\cite{bnl}}.
The results demonstrate a good agreement between theory and
experiment (see  Fig.3).
The lowest collective $\gamma$ vibrational frequency
for the positive signature states (even spins)
becomes zero at $\hbar\Omega_{\rm cr} \sim 0.324$ and $0.33$ MeV for $^{156}$Dy 
and $^{158}$Er, respectively.
As discussed  above, near the rotational
frequency $\Omega_{\rm cr}$ the backbending occurs in the considered cases.

\begin{figure}[ht]
\includegraphics[height = 0.3\textheight]{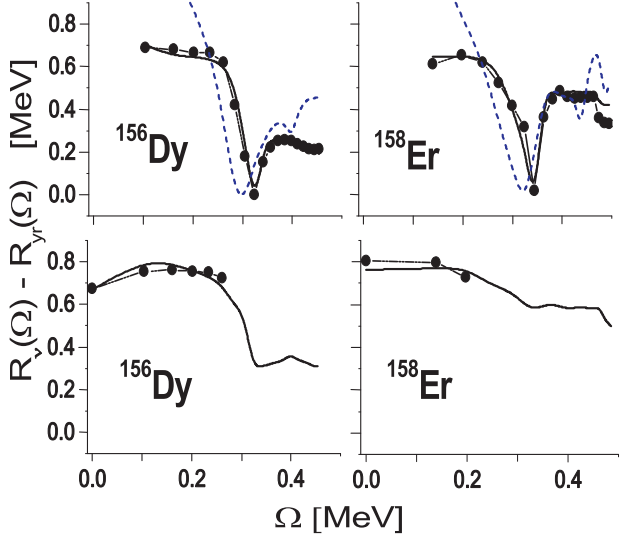}
\caption{(Color online) The excitation energies in the rotating frame
$\omega_{\nu}(\Omega)$
for  the positive signature $\gamma$ (top panel) and $\beta$ vibrational
(bottom panel) states. The results of calculations are connected by a
solid line.
The lowest two-quasiparticle states shown at the upper panel are
connected by a dash line. The experimental values for different
excitations are indicated by open circles connected by a thin line to
guide eyes.}
\label{fig3}
\end{figure}

In order to understand this correlation, let us consider an
axially deformed system, defined by the Hamiltonian ${\tilde H}$
in the laboratory frame, that rotates about a symmetry axis z with
a rotational frequency $\Omega$. The angular momentum is a good
quantum number and, consequently,
$[\hat{J}_z,O_K^{\dagger}]=KO_K^{\dagger }$. Here,  the CRPA
phonon  $O_K^{\dagger}$ describes the vibrational state with $K$
being the value of the angular momentum carried by the phonons
$O_K^{\dagger}$ along the symmetry axis, $z$ axis. Thus, one
obtains \beq [H_{{\Omega}},O_K^{\dagger}]= [{\tilde H} -\Omega
\hat{J}_z, O_K^{\dagger }]= ({\tilde \omega}_K- K\Omega)
O_K^{\dagger } \equiv \omega_K O_K^{\dagger }. \label{man} \eeq
This equation implies that at the rotational frequency
$\Omega_{\it cr}={\tilde \omega}_K/K$ one of the CRPA frequency
$\omega_K$ vanishes in the rotating frame (see discussion in
Refs.\onlinecite{M96,HN}). At this frequency we could expect the
SSB effect of the rotating mean field due to the appearance of the
Goldstone boson related to the multipole-multipole forces with
quantum number $K$. For an axially deformed system
one obtains the breaking of the axial symmetry, since the lowest
critical frequency corresponds to $\gamma$ vibrations with $K=2$
\cite{M96,HN}. In contrast to nonrotating case \cite{R},
the RPA {\it does not} break down at the bifurcation point, since
in the laboratory system the corresponding vibration still persists.

\begin{figure}[ht]
\includegraphics[height = 0.2\textheight]{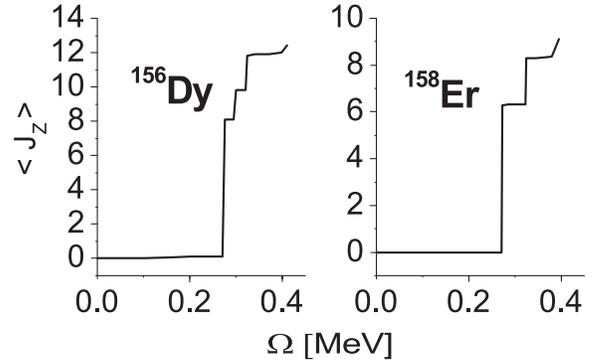}
\caption{The rotational dependence of the
mean field value $\langle J_z \rangle$ (in units $\hbar$)
for $^{156}$Dy (left panel)
and $^{158}$Er (right panel).
The noncollective angular momentum is built
by the alignment of the separate
two-quasiparticle pairs till $\Omega_{\rm cr}\approx 0.32$ (MeV) in
both nuclei.}
\label{fig4}
\end{figure}

Guided by this analysis, we solve the HB equations
with rotation about the symmetry axis z (the noncollective regime).
Since the angular momentum of a quasiparticle state
is conserved, the lowest two-quasiparticle configuration with
the largest  deformation aligned orbitals
builds the total angular momentum of the system
when it crosses the ground state configuration with
zero angular momentum.
The total angular momentum increases each time
stepwise by an amount of angular momentum
carried by a two-quasiparticle configuration that crosses
the renewed ground state configuration.
For $^{156}$Dy (see Fig.\ref{fig4}) $\langle J_z \rangle$ is zero
for all rotational frequencies in the range of values
$0\leq\hbar\Omega\leq0.275$ MeV.
At $\hbar\Omega=0.275$ MeV the energy of
the neutron two-quasiparticle state originating from the shells
$h_{9/2}\otimes f_{7/2}$ (for $\Omega=\Delta=0$
it is the configuration built from the Nilsson states
$\frac{9}{2}[505] \otimes \frac{7}{2}[514]$)} goes to zero.
The nuclear angular momentum  is
determined by the angular momentum of this two-quasiparticle
state:
$\langle J_z \rangle = \frac{9}{2}+\frac{7}{2} = 8\hbar$. 
Note that one would expect the onset of the nonaxiallity at this frequency, according
to the quasiparticle picture. According to our results,
at $\hbar\Omega = 0.31-0.32$ MeV the expectation value $\langle J_z \rangle$
is increased stepwise by next four units. The system remains an 
axially deformed creating a state with
$\langle J_z \rangle = 12 \hbar$. At $\hbar\Omega_{\rm cr}=0.324$ MeV
the system acquires additional two units and
the shape transition occurs to a nonaxially deformed  shape.
We recall that at this frequency $\gamma$ vibrations
of the positive signature carrying two units of the angular momentum vanish.
Comparing the calculated value of the critical frequency with
the experimental estimation
$\hbar\Omega_{\rm cr}={\hbar\tilde \omega}_{K=2}/2=0.691/2 MeV \sim 0.345$ MeV,
one could find a good agreement between theory and experiment.
As was mentioned above, the potential energy surface is very shallow at large
rotational frequencies. In fact, the energy minima for a rotation around the
axis z and the axis x are almost degenerate. The difference is about 15 keV near 
the critical rotational frequency. At the bifurcation point a competition between
the collective (around the axis x) and noncollective rotations breaks the
axial symmetry and leads to nonaxial shapes.

For $^{158}$Er the experimental critical
value is expected  at
$\hbar\Omega_{\rm cr}={\hbar\tilde \omega}_{K=2}/K=
0.613/2$ MeV $\sim 0.306$ MeV.
The angular momentum is zero up
to the rotational frequency $\hbar\Omega=0.272$ MeV.
At this frequency the proton two-quasiparticle state originating
from $g_{7/2}\otimes d_{5/2}$ shells
(for $\Omega=\Delta=0$ this configuration is built from the Nilsson states
$\frac{7}{2}[404]\otimes \frac{5}{2}[402]$) 
contributes to the value of the angular momentum
$\frac{7}{2}+\frac{5}{2} = 6\hbar$.
The system remains the axially deformed,
getting another two units of the angular momentum,  up to
$\hbar\Omega_{{\rm cr}}\approx 0.33$ MeV, where
$\gamma$ vibrations of the positive signature
are vanished in the rotating frame. 
Again, there is a tiny energy 
difference between the collective and noncollective rotations $(\sim 15 keV)$.
At this frequency the expectation value $\langle J_z \rangle$ is increased
by two more units and for $\Omega >\Omega_{{\rm cr}}$ the system is driven into
the domain of triaxial shape. 

According to the CRPA \cite{KN}, the reduced
transition probability  between
$\gamma$ (one-phonon) states of the positive signature
$B(E2, \Delta I=2) \sim |\langle Q_2^{\sigma=+}(\Delta I=2)\rangle|^2$ is
of the same order of magnitude as the collective electric quadrupole transitions
between yrast (vacuum) states. In fact, in both nuclei the $B(E2,\Delta I=2)$
strength exceeds few tens of Weisskopf units. Detailed calculations
of the transition probabilities are beyond
the scope of the present paper and will be presented elsewhere.

Summarizing, for the first time the cranking HB and RPA equations
are solved in a self-consistent manner. We obtain a good agreement
with available experimental data
for low-lying vibrational excitations at high spins
in $^{156}$Dy and $^{158}$Er.
According to our analysis,
the alignment decreases the pairing correlations and, consequently,
the $\gamma$ vibrations are softening in axially deformed nuclei.
We stress that the two-quasiparticle alignment does not create the backbending.
Rotating around the axis which is
perpendicular to the symmetry axis, we found 
that in $^{156}$Dy and $^{158}$Er
the backbending occurs at the critical rotational frequency
$\Omega_{cr}\approx{\tilde\omega}_{K=2}/2$
where ${\tilde\omega}_{K=2}$ is a $\gamma$ vibrational excitations in
the laboratory frame at $\Omega=0$.
At this frequency the positive signature $\gamma$ vibrational 
excitations vanish in the rotating frame, i.e., $\omega_{K=2}=0$.
As a result, the nuclear mean field spontaneously
breaks the axial symmetry and gives rise to nonaxially deformed
shape in the rotating frame. 
Thus, the inclusion of quantum fluctuations around the mean field approach 
extending our understanding of the backbending phenomenon.

\section*{Acknowledgments}
We thank S. Frauendorf for fruitful discussions and
are grateful to L. Kaptari for valuable suggestions.
This work was partly supported by the Czech grant agency under the contract
No. 202/02/0939  and by Grant No.\ BFM2002-03241
from DGI (Spain). R. G. N. gratefully acknowledges support from the
Ram\'on y Cajal programme (Spain).

\end{document}